\documentclass[aps,prb,twocolumn,superscriptaddress,showpacs]{revtex4-1}
\usepackage{amsmath}
\usepackage{color}
\usepackage{graphicx}
\usepackage{epsfig}
\usepackage{amsfonts}

\begin{document}

\title{Phases, collective modes, and non-equilibrium dynamics of dissipative Rydberg atoms}

\author{S. Ray}
\affiliation{Indian Institute of Science Education and
Research-Kolkata, Mohanpur, Nadia-741246, India}
\author{S. Sinha}
\affiliation{Indian Institute of Science Education and
Research-Kolkata, Mohanpur, Nadia-741246, India}
\author{K.~Sengupta}
\affiliation{Department of Theoretical Physics, Indian Association
for the Cultivation of Science, Jadavpur, Kolkata-700032, India.}

\date{\today}

\begin{abstract}

We use a density matrix formalism to study the equilibrium phases
and non-equilibrium dynamics of a system of dissipative Rydberg
atoms in an optical lattice within mean-field theory. We provide
equations for the fixed points of the density matrix evolution for
atoms with infinite on-site repulsion and analyze these equations to
obtain their Mott insulator- superfluid (MI-SF) phase boundary. A
stability analysis around these fixed points provides us with the
excitation spectrum of the atoms both in the MI and SF phases. We
study the nature of the MI-SF critical point in the presence of
finite dissipation of Rydberg excitations, discuss the fate of the
superfluidity of the atoms in the presence of such dissipation in
the weak-coupling limit using a coherent state representation of the
density matrix, and extend our analysis to Rydberg atoms with finite
on-site interaction via numerical solution of the density matrix
equations. Finally, we vary the boson (atom) hopping parameter $J$
and the dissipation parameter $\Gamma$ according to a linear ramp
protocol. We study the evolution of entropy of the system following
such a ramp and show that the deviation of the entropy from its
steady state value for the latter protocol exhibits power-law
behavior as a function of the ramp time. We discuss experiments
which can test our theory.

\end{abstract}

\pacs{32.80.Ee, 03.75.Lm, 05.30.Jp, 05.30.Rt, 64.60.Ht}

\maketitle

\section{Introduction}
In recent years the experimental and theoretical study of Rydberg
atoms has become an active area of research in the field of
ultracold quantum gases and condensed matter
physics\cite{ryd1,ryd2,ryd3,rydspin,ryd4,ryd5,ryd6,ryd7,ryd8,ryd9,ryd10,ryd11,ryd12}.
In Rydberg atoms, a valence electron may be excited to a state of
large principal quantum number by suitable laser driving. The
presence of large polarizibility of such excited states leads to
strong Van der Walls force between these excited atoms; this force
turns out to be the main ingredient for strong correlation in
ultracold Rydberg system. Such strong Van der Walls interaction of a
Rydberg excited atom hinders another excitation within a certain
radius from it. This phenomenon, known as Rydberg (dipole) blockade,
has been studied both theoretically \cite{ryd1} and experimentally
\cite{ryd2}. It has been shown that the long range Van der Walls
interaction between the Rydberg excited atoms leads to the formation
of density ordering and translational symmetry broken phases like
supersolid (SS) droplets within superfluid (SF) phase of these
atoms\cite{ryd3}. Also an effective spin model with
antiferromagnetic ordering can be realized in a system of frozen
Rydberg atoms (with vanishing kinetic energy of the atoms) due to
the long range interaction\cite{rydspin}. In a recent experiment on
Rydberg atoms in an optical lattice it has been demonstrated that
the superfluidity is not destroyed by the strong interaction between
the excited atoms\cite{ryd12}. Various exotic phases and collective
excitations of Rydberg gas in an optical lattice has been studied
theoretically\cite{saha}.

Another important property of a Rydberg atom is its finite lifetime
in the excited state \cite{ryd_review}. Due to such a finite decay
rate of the excitations, ultracold Rydberg gases constitute a
non-equilibrium quantum many body system. In a recent experiment an
ensemble of dilute  thermal Rydberg gas has been shown to undergo a
non-equilibrium phase transition between states of high and low
Rydberg occupancy which is induced by resonant dipolar interaction
between the atoms \cite{carr}. Such a transition is marked by
critical slowing down which can be detected by measurement of time
taken by the system to reach the steady after being optically
excited by an external laser source\cite{carr}. This phenomenon has
been theoretically analyzed by modeling the frozen Rydberg atoms as
a dissipative spin system with decay and
dephasing\cite{lesanovsky1}. Such an analysis revealed the existence
of a non-conserved order parameter with a definite $Z_2$ symmetry
which attains finite values as the system moves across the
non-equilibrium transition to the large dipolar interaction regime.
Further, such frozen Rydberg atoms, in the presence of a bipartite
lattice and in the zero hopping limit, was shown to undergo a
quantum phase transition from an uniform to a translational symmetry
broken (antiferromagnetic) state \cite{cross}; the details of the
transition has been analyzed using density matrix formulation within
mean-field theory which indicated the existence of non-equilibrium
steady states in these systems \cite{cross}. Various other
non-equilibrium properties of frozen Rydberg gases and their
relaxation dynamics have been studied theoretically
\cite{lesanovsky2, clark}. However the fate of the superfluid phase
and the Mott-superfluid transition of these atoms remains beyond the
scope of the models used in these studies since the frozen limit of
the Rydberg atoms is analogous to setting their hopping strength $J$
to zero in a lattice. Thus the effect of finite lifetime of the
Rydberg excitations on the superfluid order parameter and the Mott
insulator-superfluid (MI-SF) transition of such systems has not been
studied so far. However, recent experiments has led to realization
of Bose-condensate of Rydberg atoms \cite{ryd12}. The interpretation
of such experimental results and its analog in the strong coupling
regime clearly necessitates a formalism for treating Rydberg atoms
in the presence of a finite excitation lifetime beyond the frozen
limit; to the best of knowledge such a formalism has not been
developed so far and we  aim to fill up this gap in the present
work.

In  this work, we analyze the Rydberg atoms using a density matrix
formalism and within mean-field theory. We chart out the possible
homogeneous steady states of Rydberg atoms in an optical lattice in
the strong coupling regime with a finite hopping strength $J$ and
allowing for a finite decay time for Rydberg excitations
characterized by a decay rate $\Gamma$, as a function of $J$ and
$\Gamma$. We analyze the time evolution of the Rydberg bosons and
obtain the possible steady states of the system from the fixed
points of the density matrix evolution equations. Our analysis of
the fixed point equations, in the limit of infinite on-site boson
repulsion (hardcore limit), leads to determination of the MI-SF
phase boundary of these atoms as a function of $J$ and $\Gamma$. We
also carry out a stability analysis around these fixed points to
chart out the collective modes of the atoms in each of the phases.
We supplement our analytical results with numerical studies for
bosons with finite on-site repulsion and show that the phase diagram
so obtained is qualitatively similar to that for atoms with infinite
on-site repulsion for a wide range of chemical potential. Next, we
use a coherent state representation of the boson density matrix deep
in the SF phase in the weak-coupling limit and obtain their
collective modes; in particular, we chart out the variation of the
group velocity of the bosons with $\Gamma$. Finally, we study
non-equilibrium dynamics of these bosons subjected to a linear ramp
across the transition point and discuss entropy production which
accompanies such ramps. We point out that the entropy of the atoms
during such ramp dynamics shows qualitatively distinct behavior
depending on whether $J$ or $\Gamma$ is tuned to cross the MI-SF
phase boundary. We discuss experiments that may test our theory. We
note that our work, being carried out using a mean-field formalism,
is expected to be qualitatively accurate for two- and
three-dimensional Rydberg systems. In addition, to the best of our
knowledge, it provides the first theoretical analysis of
superfluidity, MI-SF transition, collective modes and
non-equilibrium dynamics of such dissipative Rydberg atoms in the
strong-coupling regime; we therefore expect this work to be of
interest to both experimentalists and theorists working on this
system.

The plan of the rest of the work is as follows. In Sec.\
\ref{hcore}, we formulate the evolution equation for the density
matrix, obtain analytical expressions of the MI and SF fixed points
from it in the infinite on-site repulsion (hard core) limit, and
chart out the MI-SF phase boundary using these equations. This is
followed by Sec.\ \ref{ptcm}, where we carry out the stability
analysis around these fixed points and obtain the collective modes
in each of the phases. We also carry out numerical analysis of the
phases of the soft-core bosons in this section. This is followed by
Sec.\ \ref{deepsf}, where we use a coherent state representation of
the density matrix to discuss the fate of superfluidity of the
Rydberg bosons in the weak-coupling limit. Next, in Sec.\
\ref{neqd}, we discuss non-equilibrium dynamics of these systems for
a linear ramp protocol of the hopping strength $J$ and the
dissipation constant $\Gamma$ and contrast several features of such
dynamics from their dissipationless counterpart in closed quantum
systems. Finally, we summarize our results, discuss possible
experiments relevant to our theory,  and conclude in Sec.\
\ref{dis}.

\section{Phases of Hardcore boson}
\label{hcore}

The Hamiltonian of the Rydberg atoms in the presence of an optical
lattice are described by $H=H_0+H_1+H_2$ with
\begin{eqnarray}
H_0 &=&\Omega\sum_i(a_i^\dagger b_i+h.c.)-\mu_0\sum_i\hat{n}_i+\Delta\sum_in_i^b \nonumber\\
&& + \frac{U}{2}\sum_i\hat{n}_i^a(\hat{n}_i^a-1) +\lambda_0 U\sum_i\hat{n}_i^a\hat{n}_i^b \label{On site term}\\
H_1&=&-\frac{J}{2}\sum_{\langle ij\rangle}(a_i^\dagger a_j+\eta b_i^\dagger b_j+h.c.) \label{Hopping term}\\
H_2&=&\frac{V_d}{2}\sum_{ij}\frac{\hat{n}_i^b\hat{n}_j^b}{\vert
r_i-r_j\vert ^6} \label{Interaction term}
\end{eqnarray}
where $a_i(b_i)$ are annihilation operator for boson at site $i$ in
the ground(excited) state, ${\hat n}_i^{a(b)}= a_i^{\dagger} a_i
(b_i^{\dagger} b_i)$ are the corresponding number operators with
${\hat n}_i= \hat n_i^a + {\hat n}_i^b$, $V_d$ is the interaction
strength between neighboring bosons in the excited states,
J($\eta$J) is the hopping strength of ground(excited) state bosons,
$\mu_0$ is the chemical potential, $U (\lambda_0 U)$ are the on-site
interaction strength between two bosons in ground (different)
states, $\Omega$ is the effective Rabi coupling between Rydberg
bosons in the ground and the excited state, and $\Delta$ is the
on-site energy cost for producing a Rydberg excitation which can be
tuned to be negative in experimental systems. In the rest of this
work we shall set $\hbar =1$ and measure all energy (time) in units
of $\Omega(1/\Omega)$ unless stated otherwise. The dynamics of the
system's density matrix $\hat{\rho}$ is governed by
\begin{eqnarray}
\frac{\partial \hat{\rho}_i}{\partial t}&=&
-i\left[\hat{H}^{MF}_i,\hat{\rho}_i\right]+ {\hat {\mathcal L}}
\hat \rho \nonumber\\
{\hat {\mathcal L}} \hat \rho  &=& \sum
_{j}(\hat{L}_{ij}\hat{\rho}_i
\hat{L}_{ij}^{\dagger}-\frac{1}{2}\hat{L}_{ij}^{\dagger}\hat{L}_{ij}\hat{\rho}_i-\frac{1}{2}\hat{\rho}_i\hat{L}_{ij}^{\dagger}\hat{L}_{ij})
\label{mean field QME}
\end{eqnarray}
where $L=\sqrt{\Gamma} a^{\dagger}b$ represents the decay of an
Rydberg atom from the excited to the ground state. In this section,
we shall study these equations for hardcore bosons for which $U \to
\infty$. For such bosons, the states of the systems can be expressed
in the basis $|n_a,n_b\rangle$; at each site the boson states are
linear combination of $|0,0\rangle$, $|1,0\rangle$ and
$|0,1\rangle$. Further within mean-field theory and as long as we
restrict ourselves to the study of the homogeneous phases of the
system, the Hamiltonian of the system can be written as $H_{\rm mf}=
\sum_i H_i$ where $H_i$ is given by
\begin{equation}
H_i = \left( \begin{array}{ccc}
0 & -\frac{Jz}{2}\langle a_j^{\dagger}\rangle & -\frac{\eta Jz}{2}\langle b_j^{\dagger}\rangle \\
\frac{-Jz}{2}\langle a_j\rangle & -\mu_0 & 1 \\
-\frac{\eta Jz}{2}\langle b_j\rangle & 1 &
-\mu_0+\Delta+\frac{Vz}{2}\langle n_{2j}\rangle
\end{array}\right)
\label{HMF}
\end{equation}
where $j$ denote any one of the nearest-neighbor sites of $i$ and
$z$ is the coordination number of the lattice. We note here that for
homogeneous phases $\langle O_j \rangle \equiv \langle O \rangle$
for any operator $O$. Also, all energies in Eq.\ \ref{HMF} are
dimensionless quantities scaled in units of $\Omega$; for example $J
\equiv J/\Omega$. The density matrix can be written using the
above-mentioned basis, as
\begin{equation}
\hat{\rho}_i = \left( \begin{array}{ccc}
n_0 & \alpha & \beta \\
\alpha^* & n_1 & \gamma \\
\beta^* & \gamma^* & n_2
\end{array}\right)
\label{denmat}
\end{equation}
where $\alpha = \langle a^{\dagger}\rangle$ and $\beta = \langle
b^{\dagger}\rangle$ are the order parameters for ground and excited
states state superfluidity, $\gamma = \langle a^{\dagger}b\rangle$,
$n_i = \langle \hat n_i\rangle$ for $i=1,2$ are the boson number
density for ground and excited states, and $n_0 = \langle
|0,0\rangle \langle 0,0|\rangle$. Here the average of the operators
are represented by $\langle . \rangle = {\rm Tr} [.
\hat{\rho}_{i}]$.
Using Eqs.\ \ref{mean field QME}, \ref{HMF} and \ref{denmat}, we get
the following dynamical equations
\begin{subequations}
\begin{eqnarray}
\dot{n}_0 &=& 0 \quad \dot{n}_{1(2)} = -(+)i (\gamma
^{*}-\gamma)+(-)\Gamma n_2 \label{eq7b}\\
\dot{\alpha} &=& i\left[(z J(n_1-n_0)/2 - \mu _{0})\alpha
+ (\eta z J \gamma ^{*}/2 + 1) \beta \right] \label{eq7c} \\
\dot{\beta} &=& i\left[(z J \gamma/2 +1)\alpha + (\eta
zJ (n_2 - n_0)/2-\mu_0 \right.\nonumber\\
&& \left.+ \Delta + zV n_2/2)\beta \right] -
\Gamma\beta/2 \label{eq7d} \\
\dot{\gamma} &=& i\left[Jz \alpha ^{*}\beta(1-\eta)/2 -
(n_2 - n_1) \right. \nonumber\\
&& \left. + (\Delta + z V n_2/2)\gamma \right] - \Gamma \gamma/2
\label{eq7e}
\end{eqnarray}
\label{eq7}
\end{subequations}
where dot represents the derivative with respect to scaled time
$t\equiv t\Omega$. From the single site mean field dynamical
equations we see that $n_0$ is a free parameter which remains
constant with time. As a result, the density of the system remains
constant in the dynamics and is given by $\langle \hat{n}\rangle = 1
-n_0$ due to the constraint $n_0 + n_1 + n_2=1$. For the time
evolution of the density matrix the initial conditions to be
supplied for solution of Eq.\ \ref{eq7}. To analyze the long time
dynamics and steady states numerically, we are going to choose the
initial density matrix as a pure state constructed from the ground
state of the Gutzwiller wavefunction for $H_{\rm mf}$, since it
represents the physical steady state with vanishing entropy for
$\Gamma=0$. Also this choice of initial density matrix is most
natural in an experimental setup. In what follows, we first look
into the fixed point structure of Eq.\ \ref{eq7}. We find that there
are two distinct fixed points; the first corresponds to the MI phase
and has $|\alpha|=|\beta|=0$ while the second corresponds to the SF
phase with non-vanishing $\alpha$ and $\beta$.

{\it MI phase}: For the Mott fixed point, $\gamma_R$ and $n_2$ can
be obtained from Eqs.\ \ref{eq7b} and \ref{eq7e} after setting
$|\alpha|=|\beta|=0$ in Eqs.\ \ref{eq7c} and \ref{eq7d}. Such a
procedure yields
\begin{eqnarray}
&\gamma_R = -\tilde{V} n_2, \quad \gamma_I= {\tilde \Gamma} n_2
\label{gamaRmot}\\
&(n_0+2n_2-1)+\tilde{V}^2n_2 + \tilde{\Gamma}^2n_2 = 0,
\label{n2mot}
\end{eqnarray}
where $\tilde{V}=(2\Delta +Vz n_2)/2$ and $\tilde{\Gamma}=\Gamma/2$.
We note from the expression of $\tilde V$ that Eq.\ \ref{n2mot}
constitutes a cubic equation for $n_2$. From Eq.\ \ref{gamaRmot} and
\ref{n2mot}, we find that there are two distinct MI phases. The
first corresponds to $n_2=0$ and $n_0=1$ which is the boson vacuum;
such a MI phase occurs for large negative $\mu_0$. In this case,
there is no time evolution and $\dot \rho = 0$ for all $t$. The
second corresponds to $n_2,n_1 \ne 0$ and $\gamma \ne 0$ as
determined by Eqs.\ \ref{gamaRmot} and \ref{n2mot}; this constitutes
a MI phase with finite boson density. We concentrate on the latter
phase for the rest of this work.

{\it SF phase}: Next, we look for the fixed point with non-zero
$\alpha = |\alpha| \exp(i \theta_1)$ and $\beta = |\beta| \exp(i
\theta_2)$ which corresponds to the steady state in the SF phase. We
note at the outset that while $|\alpha|$ and $|\beta|$ assumes
constant values in the steady state, $\alpha$ and $\beta$ need not
be constant since the global phase of the order parameter may
fluctuate. The relative phase $\theta_r = \theta_2-\theta_1$ of
these superfluid order parameters also attain a constant value in
the steady state. To find the fixed point values of $\alpha$ and
$\beta$, it is therefore convenient to rewrite Eq.\ \ref{eq7} as
\begin{subequations}
\begin{eqnarray}
|\dot \alpha |&=& \eta z J|\beta|(\gamma_I\cos\theta_{r}-\gamma_R\sin\theta_{r})/2-|\beta|\sin\theta_{r} \label{eq10a} \\
|\dot \beta | &=& z J\left(\gamma_{R}\sin\theta_{r}-\gamma_{I}\cos\theta_{r}\right)|\alpha|/2 +|\alpha|\sin\theta_{r} \nonumber\\
&& - {\tilde \Gamma}|\beta| \label{eq10b} \\
\dot{\theta}_{r} &=& \frac{Jz}{2} \Big[ \eta n_2 -n_1 +(1-\eta)n_0
+ (|\alpha|^2-\eta |\beta|^2) \nonumber\\
&& \times (\gamma_R \cos \theta_r + \gamma_I \sin
\theta_r)/(|\alpha||\beta|) \Big] \nonumber\\
&& + (|\alpha|^2-|\beta|^2) \cos\theta_r /(|\alpha||\beta|)  +
\tilde V \label{eq10c}
\end{eqnarray}
\label{eq10}
\end{subequations}
 Since we look for time independent solutions for
$|\alpha |$ and $|\beta |$, we set $|\dot \alpha |=|\dot \beta|=0$.
For the SF phase, defining $r=|\alpha|/|\beta|$,  we find from Eqs.\
\ref{eq10a} and \ref{eq10b}
\begin{eqnarray}
\tan\theta_{r}&=& \frac{\eta {\tilde J}\gamma_I}{1+ \eta {\tilde
J}\gamma_R} \,\, r=
\frac{\tilde{\Gamma}\sec\theta_{r}}{\tan\theta_{r}-\tilde{J}(\gamma_I-\gamma_R\tan\theta_{r})}
\label{r}
\end{eqnarray}
where $\tilde{J}=Jz/2$. Moreover, since $\dot \theta_r=0$ using Eq.\
\ref{eq10c}, we obtain
\begin{eqnarray}
&&\tilde{J}(1-2n_0-n_2)\sec\theta_{r}+\left(r^{-1}-r\right)
+\tilde{J}(\gamma_R +
\gamma_I\tan\theta_{r})\nonumber\\
&& \times \left(\eta r^{-1}-r\right) - (\eta
\tilde{J}(n_2-n_0)+\tilde{V})\sec\theta_{r}=0 \label{dth1mnsdth2}
\end{eqnarray}
Finally setting $\dot{\gamma}=0$ in Eq.\ \ref{eq7e} we obtain the
steady state values of $\gamma_R$ and $\gamma_I$ to be
\begin{eqnarray}
&& \tilde{J}(1-\eta)|\alpha ||\beta |\sin \theta_{r}+\tilde{V}\gamma_I +\tilde{\Gamma}\gamma_R = 0 \label{gammaeq}\\
&& \tilde{J}(1-\eta)|\alpha ||\beta |\cos
\theta_{r}-(n_0+2n_2-1)+\tilde{V}\gamma_R - \tilde{\Gamma}\gamma_I =
0 \nonumber\
\end{eqnarray}
We note that in the SF phase the fixed point equations assume a
particularly simple for $\eta=0$ leading to
\begin{eqnarray}
&\theta_{r} = n\pi, \quad r=1/(\tilde{J}n_2), \quad \gamma_R = -\tilde{V}n_2 \nonumber\\
&\tilde{J}^2n_2|\alpha|^2+(2n_2+n_0-1)+\tilde{V}^2n_2+\tilde{\Gamma}^2n_2=0
\label{n2alphaeq}
\end{eqnarray}
Further, in this limit, Eq.\ \ref{dth1mnsdth2} can also be
simplified to obtain a solution for $n_2$ which is given by
\begin{eqnarray}
n_2 &=&\left[4z(J+V)\right]^{-1}\left[(1-2n_0)Jz -4 \Delta
\right.\nonumber \\
&&\left.+\sqrt{((1-2n_0)Jz-4\Delta)^2+32(1+V/J)}\right]
 \label{solnn2}
\end{eqnarray}

\begin{figure}[ht]
\begin{center}
\includegraphics[scale=0.5]{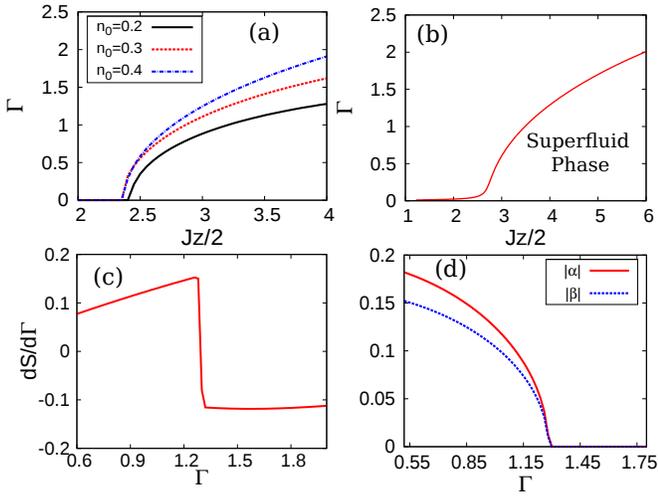}
\end{center}
\caption{ Mean field phase diagram for the MI-SF transition of the
Rydberg atom in the hardcore boson limit. (a) The phase boundary for
different density $1-n_0$ with $\eta=0$, $zV=2$, and $\mu=\Delta=0$.
(b) Phase boundary for finite $\eta=0.2$; all other parameters are
same as (a). (c) Plot of the entropy derivative $dS/d\Gamma$ as a
function of $\Gamma $ for $zJ=8$; all other parameters are same as
in (b). (d) Plot of the order parameter amplitudes $|\alpha|$ and
$|\beta|$ as a function of $\Gamma$; all parameters are same in (c).
Note that $dS/d\Gamma$ shows a jump and the order parameter
amplitudes vanish at the phase boundary.} \label{PhDiag1}
\end{figure}

Eqs. \ref{eq7}..\ref{solnn2} constitute the complete structure of
fixed points of the system both for the MI and SF phases. It turns
out that only one of the two fixed points obtained is stable; their
change of stability occurs at the phase transition between the MI
and the SF phases. This stability analysis, which provides us
information regarding the phase diagram of the system, is carried
out in details in the next section. However, we note that there is
an alternative way of obtaining the phase diagram from the fixed
point equations. For this, one solves Eqs.\ \ref{gamaRmot} and
\ref{n2mot} to get values of $n_2$, $\gamma_R$ and $\gamma_I$; these
are then substituted in Eqs.\ \ref{r} and \ref{dth1mnsdth2}. A
numerical solution of Eqs.\ \ref{r} and \ref{dth1mnsdth2} with these
substituted values of $n_2$, $\gamma_R$ and $\gamma_I$ yields a
relation between $J$ and $\Gamma$. Since this relation holds, by
construction, for both the MI and SF phases, it represents the MI-SF
phase boundary. The SF phase is stabilized for $\Gamma(J)$
lower(higher) than their respective values on the phase boundary;
the opposite holds for the MI phase. We note that the phase diagram
obtained in this manner coincides with that obtained via stability
analysis of the fixed points outlined in the next section.

\begin{figure}[ht]
\begin{center}
\includegraphics[scale=0.48]{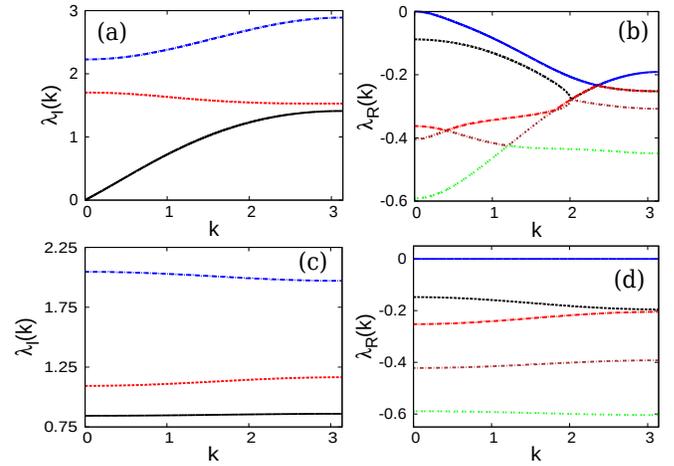}
\caption{Plot indicating the stability of the fixed point and
corresponding to the collective models for the SF and MI phases. (a)
Plot of $\lambda_I(k)$ as a function $k$ (along a diagonal cut in
the Brillouin zone) for the SF phase showing gapless and gapped
excitations. The parameters used are for $zJ=8$, $\Gamma=0.8$,
$\eta=0.2$, $\Delta=\mu=0 $, and $zV=2$. (b) Plot of $\lambda_R$ as
a function of $k$ showing stability of the SF phase. All parameters
are same as in (a). (c)-(d): Similar plots as in (a) and (b)
respectively corresponding to the MI phase for $zJ=1$; all other
parameters are same as in (a). Note the absence of gapless
collective modes in the MI phase.} \label{fig4}
\end{center}
\end{figure}

The MI-SF phase diagram obtained in this manner is shown in Fig.\
\ref{PhDiag1}(a) for several representative values of $n_0$ for
$\eta=0$. A similar diagram, obtained using numerical solution of
Eqs.\ \ref{r} and \ref{dth1mnsdth2}, is shown for $\eta=0.2$. The
order of this transition can be obtained by computing the entropy
$S=-{\rm Tr}(\rho \log \rho)$ and $dS/d\Gamma$ across the
transition. We find, as shown in Fig.\ \ref{PhDiag1}(c), $
dS/d\Gamma$ shows a jump at the phase boundary; this, together with
the fact that $S$ varies smoothly across the MI-SF boundary
displaying a peak at the transition point, indicates that the
transition is second order. The vanishing of the superfluid order
parameter amplitudes $\alpha$ and $\beta$ across the transition is
shown in Fig.\ \ref{PhDiag1}(d). Thus our fixed point analysis
provides the complete phase diagram for the MI-SF transition.

\section{Stability analysis and collective modes}
\label{ptcm}

In this section, we carry out a stability analysis of the fixed
points which also enables us to obtain the dispersion of the
low-energy excitations in the MI and SF phases. This is done,
following standard prescription, as follows. For all of the elements
of the density matrix, we study small fluctuations about the steady
state values: $\rho^{ab}= \rho^{ab;(0)} + \delta \rho^{ab}_i$, where
$\rho^{ab;(0)}$ is the steady state value of $\rho^{ab}$ (Eq.\
\ref{denmat}), $\delta \rho_i^{ab}$ denotes fluctuations around the
steady state value, $i$ is the site index, and the indices $a$ and
$b$ take values from $1$ to $3$. We note that $\rho^{12} \equiv
\alpha$ and $\rho^{13} \equiv \beta$ shows oscillatory behavior at
the fixed point whereas $|\alpha|$ and $|\beta|$ take fixed values;
this allows us to parameterize $\alpha = \alpha_1 \exp(i E_f t)$ and
$\beta = \beta_1 \exp(i E_f t)$ where $E_f$ is the frequency of the
oscillation. Also we note that $\delta \rho_i^{12}$ and $\delta
\rho_i^{13}$ denote fluctuations corresponding to $\alpha_1$ and
$\beta_1$.

We substitute this form of $\rho_{ij}$ in Eq.\ \ref{eq7} and seek a
solution of the form $\delta \rho^{ab}_{i}(t)= \exp(\lambda t) \sum
_{\vec k} \exp(i\vec{k}.\vec{R_i})\delta \rho^{ab}_{\vec k}$
retaining terms which are linear order in $\delta \rho_{\vec
k}^{ab}$. We also note that since phases $\theta_{\alpha}$ and
$\theta_{\beta}$ of the SF order parameters $\alpha$ and $\beta$
respectively remain undetermined up to a global phase, we choose the
steady state value of $\beta$ to be $\beta_1= \alpha^* \beta
/|\alpha|$ so as to have a specific relative phase, where
$\alpha_1=|\alpha| \exp[i \theta_{\alpha}]$ is the steady state
value of $\alpha$. This leads us to
\begin{eqnarray}
\lambda \delta n_{1\vec k} &=& -i\left[(\delta \gamma_{\vec k}^*-\delta \gamma_{\vec k})-J(\alpha_1^*\delta\alpha_{\vec k}-\alpha_1\delta\alpha_{\vec k}^*)\right. \nonumber\\
&& \left.\times (z-\epsilon(\vec k))/2\right]+\Gamma\delta n_{2 \vec k} \label{deltan1}\\
\lambda \delta n_{2\vec k} &=& i\left[(\delta \gamma_{\vec k}^*-\delta \gamma_{\vec k})+ \eta J (\beta_1^*\delta \beta_{\vec k}-\beta_1\delta \beta_{\vec k}^*)\right. \nonumber\\
&& \left. \times (z-\epsilon(\vec k))/2\right]-\Gamma \delta n_{2\vec k} \label{deltan2} \\
\lambda \delta \alpha_{\vec k} &=& i\left[z J \alpha_1(\delta n_{1\vec k}+\delta n_{2\vec k}/2)+\left(J(2n_1+n_2-1)\epsilon(\vec k) \right. \right.\nonumber\\
&& \left. \left. -2\mu -2 E_f\right)\delta \alpha_{\vec k}/2 + (1+ \eta \epsilon(\vec k) J \gamma ^*/2)\delta \beta_{\vec k} \right. \nonumber\\
&& \left. + \eta z J \beta_1\delta \gamma_{\vec k}^*/2 \right]
\label{deltaalpha} \\
\lambda \delta \beta_{\vec k} &=& i\left[zJ\alpha_1\delta\gamma_{\vec k}/2 + (1 + J\gamma \epsilon(\vec k)/2)\delta \alpha_{\vec k} \right. \nonumber \\
&& \left. - (\mu + E_f+ \eta J(1-2n_2-n_1)\epsilon(\vec k)/2 \right. \nonumber\\
&& \left. -(\Delta + z V n_2/2))\delta \beta_{\vec k}
+V\beta_1 \epsilon(\vec k)\delta n_{2\vec k} /2 \right. \nonumber\\
&& \left. + \eta z J\beta_1(\delta n_{2\vec k}+\delta n_{1\vec
k}/2)\right]-\Gamma\delta \beta_{\vec k}/2
\label{deltabeta} \\
\lambda \delta \gamma_{\vec k} &=& i\left[ z J \alpha_1^*\delta \beta_{\vec k}/2 +J \beta_1 \epsilon(\vec k) \delta \alpha_{\vec k}^*/2 -\delta n_{2\vec k}+ \delta n_{1\vec k} \right. \nonumber\\
&& \left. + (z V n_2/2 +\Delta) \delta \gamma_{\vec k}  + V\gamma
\epsilon(\vec k)\delta n_{2\vec k}/2
\right. \nonumber\\
&& \left. -\eta J \epsilon(\vec k)\alpha_1^*\delta \beta_{\vec k}/2
-\eta z J \beta_1 \delta \alpha_{\vec k}^*/2 \right]-\Gamma \delta
\gamma_{\vec k}/2
\end{eqnarray}
where, $n_1$ and $n_2$ denote the steady state values, $\epsilon(k)=
2\sum_{i=1..d} \cos(k_i)$ where $d$ is the dimension of the system,
$\delta \alpha_{\vec k}$ and $\delta \beta_{\vec k}$ are
fluctuations of $\alpha_1$ and $\beta_1$ respectively. These
equations are supported with another three equations for $\delta
\alpha_{\vec k}^*$, $\delta \beta_{\vec k}^*$ and $\delta
\gamma_{\vec k}^*$ and the system of these eight equations are to
solved along with the constraint $n_0+n_1+n_2=1$ which shall be used
to eliminate $\delta n_0$. A numerical solutions of these equations
then yields the eigenmodes of the system which is shown in Fig.
\ref{fig4}.


We first check the stability of these fixed points from such an
analysis by inspecting the real part of the eigenvalues $\lambda$.
We find that for a given $n_0$ which is chosen from the Gutzwiller
ground state, ${\rm Max}[{\rm Re}[\lambda]]$, as computed around the
SF fixed point, attains zero value along a curve in the $J-\Gamma$
plane as shown in Fig.\ \ref{PhDiag1}(a) and (b). Similarly, ${\rm
Max}[{\rm Re}[\lambda]]$, as computed around the MI fixed point
becomes zero along the same curve. This curve represents the phase
boundary between the MI and the SF phases and coincides with that
shown in Fig.\ \ref{PhDiag1}.

\begin{figure}[ht]
\begin{center}
\includegraphics[scale=0.165]{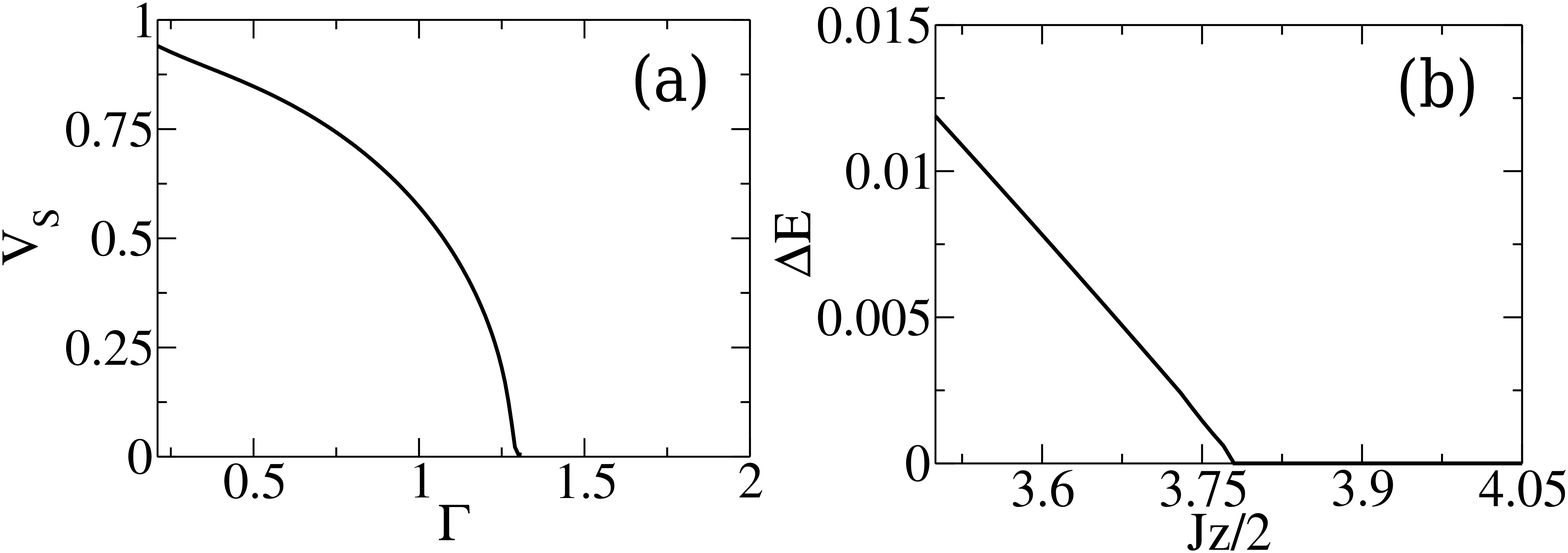}
\end{center}
\caption{(a) Plot of the superfluid velocity $v_s$ as a function of
$\Gamma$. All parameters are same as in Fig.\ \ref{fig4}(a) (b) Plot
of the $\Delta E$ as a function of $J$ for $\Gamma = 1.2$, all other
parameters are same as (a).} \label{sound_vel}
\end{figure}

Next, we concentrate on ${\rm Im}[\lambda]$ which yields the
dispersion of the eigenvalues. In the SF phase, our analysis finds
six complex eigenvalues (three pairs corresponding to the
fluctuations of $\alpha_1$, $\beta_1$ and $\gamma$) and two real
eigenvalues corresponding to $n_1$ and $n_2$. Among the three pairs
of complex eigenmodes, the imaginary part of one gives gapless sound
mode (as shown in the Fig.\ \ref{fig4}(a)) along a diagonal cut in
the Brillouin zone. This is in sharp contrast to the MI phase (Fig.\
\ref{fig4}(c) and (d))where the fluctuations corresponding to
$\alpha_1$ and $\beta_1$ get completely decoupled from $n_1$, $n_2$
and $\gamma$. In the latter sector there is one zero mode, one real
eigenvalue($\le 0$) and a pair of complex eigenvalues. Also we note
the presence of a gapless sound mode in the SF phase (top left panel
of Fig.\ \ref{fig4}) whereas no such mode exists in the insulating
phase (bottom left panel of Fig.\ \ref{fig4}). We calculate the
sound velocity($v_s=  \lim_{k\rightarrow 0} d\omega /dk \equiv
\lim_{k\rightarrow 0}  d\lambda_I/dk$) from the linear dispersion
obtained in SF phase. We note that (see Fig.\ \ref{sound_vel}) for a
fixed $J$, $v_s$ decreases with increasing $\Gamma$. We also find
that the dispersion becomes quadratic ($\omega \sim k^2$) at the
transition point (as shown in Fig.\ \ref{sound_vel}a) which
indicates a dynamical critical exponent $z=2$ for the transition.
Further, we find that the energy gap $\Delta E$ at $k=0$ satisfies
$\Delta E \sim (\Gamma -\Gamma_c)$ (see Fig.\ \ref{sound_vel}(b))
and $\Delta E \sim (J -J_c)$. Since $\Delta E \sim (J -J_c)^{z\nu}$
at the critical point, this indicates that the correlation length
exponent $\nu$ has a value $1/2$ which is in accordance with
mean-field theory.

\begin{figure}[ht]
\begin{center}
\includegraphics[scale=0.165]{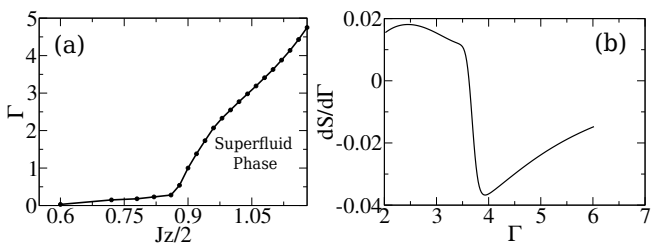}
\end{center}
\caption{(a) The MI=Sf phase boundary of soft core bosons for
$U=1.5$, $\lambda_0 = 3$, $\eta = 0.2$, $\Delta = \mu=0$, and
$zV=2$. (b) Plot of $dS/d\Gamma$ as a function of $\Gamma$ for
$zJ=2.24$; all other parameters are same as (a). Note that
$dS/d\Gamma$ exhibits sharp change in slope at the critical point.}
\label{PDsoft}
\end{figure}

Before ending this section, we note that an analogous analysis can
be carried out numerically for soft core bosons which have  finite
on-site interaction $U$. In this case, we construct the initial
density matrix of the system in the following way,
\begin{equation}
\hat{\rho} = |\psi \rangle \langle \psi |
\end{equation}
where $|\psi\rangle$ is the ground state Gutzwiller wavefunction
$\vert\psi\rangle = \Sigma_{n^a,n^b}f_{n^a,n^b}\vert
n^a,n^b\rangle$. This wavefunction can be found out numerically by
minimizing the energy functional $\langle \psi |H|\psi \rangle$. We
then numerically evolve the density matrix using the Quantum Master
Equation (QME) (Eq.\ \ref{mean field QME}) and find out the MI and
SF steady states from such evolution. A stability analysis of these
steady states, analogous to that described in the earlier part of
this section, leads to the phase diagram for the MI-SF transition as
shown in Fig.\ \ref{PDsoft}a. Similar to the hardcore boson, the
transition line (see Fig.\ \ref{PDsoft}a) corresponds to the
vanishing of the order parameters $\Delta_{a(b)} = \langle
a(b)\rangle$. We observe near the phase boundary and for small
$\Gamma$, the relaxation time rapidly increases analogues to the
critical slowing down in phase transition. Analogous to the
susceptibility of the spin systems, the derivatives of the order
parameters $\Delta_{a(b)}$ with respect to $\Gamma$ exhibit a peak
at the phase boundary which becomes sharper for longer time
evolution. Finally we calculate the derivative of entropy $S$ with
respect to $\Gamma$, which is shown in Fig.\ \ref{PDsoft}b. A jump
in the first derivative of the entropy $S$ across the phase boundary
(see Fig.\ \ref{PDsoft}b) indicates a continuous second order
dynamical transition in the system of soft-core bosons. We note that
the phase diagram obtained for bosons with finite but large $U$ is
qualitatively similar to that obtained in the $U \to \infty$ limit
in Sec.\ \ref{hcore}; this justifies our detailed analysis of the
case of hardcore bosons where it is possible to obtain analytical
results for the fixed points.

\section{Deep Superfluid limit}
\label{deepsf}

\begin{figure}[ht]
\begin{center}
\includegraphics[scale=0.17]{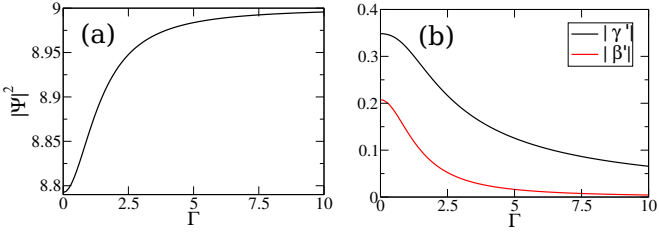}
\end{center}
\caption{(a) Plot of the steady state value of the ground state
boson density $|\psi|^2$ with $\Gamma$ showing the increase of
ground state boson density with increasing $\Gamma$. (b) Plot of
steady state values of $|\gamma^{\prime}|$ and $\beta^{\prime}$ as a function of
$\Gamma$. For both plots, $zV=2$, $zJ=1$, $\eta=\mu=1$, $U=0.6$ and
$\Gamma = 0.2$. The total boson density is fixed at
$n_{total}=9.0$.} \label{steadySF}
\end{figure}

In this section, we look into the fate of the superfluidity of the
bosons in the presence of the dissipation. To this end, we adopt the
coherent state description \cite{sudarsan} of the density matrix for
the bosons. We consider the coherent state
\begin{eqnarray}
|\psi (t), \sigma \rangle &=& e^{-|\psi (t)|^2/2}e^{\psi (t)
a^{\dagger}}|0, \sigma\rangle \label{coh1} \end{eqnarray} where $a$
is the annihilation operator for ground state boson, $\sigma$ is the
spin index where $|\uparrow\rangle(|\downarrow\rangle)$ represents
$1(0)$ boson in the excited state, and we have used the fact that
for hardcore bosons we can replace $b(b^{\dagger})$ by
$\sigma^{-}(\sigma^{+})$ where $\sigma^{-}(\sigma^{+})$ are
lowering(raising) operators. The number operator in the excited
state($\hat{n}_b$) can be replaced by $(1+\hat{\sigma}_z)/2$. The
mean-field single-site Hamiltonian can now be written as
\begin{eqnarray}
H &=& (a^{\dagger}\sigma ^{-}+\sigma ^{+}a)-zJ(\phi _aa^{\dagger}+\phi _a^{*}a)/2 \nonumber\\
&& -\eta z J(\phi _b\sigma ^{+}+\phi _b^{*}\sigma ^{-})/2 +U
\hat{n}_a(\hat{n}_a-1)/2  \nonumber\\
&& +\lambda_0 U \hat{n}_a(1+\hat{\sigma}_z)/2+ zV \langle
n_{bj}\rangle (1+\hat{\sigma}_z)/4 \nonumber\\
&& -\mu(\hat{n}_a+(1+\hat{\sigma}_z)/2)) \label{HMFDeep}
\end{eqnarray}
where $J(\eta J)$ are hopping strength for ground(excited) state
bosons, V is the interaction strength between excited states
$\phi_a=\langle a_j\rangle$, $\phi_b=\langle \sigma^{-}_j\rangle$
where j is the nearest neighbor index and other terms carry usual
meaning. The density matrix of the system can be written in the
basis of $|\psi (t),\uparrow \rangle$, $|\psi (t),\downarrow
\rangle$ as,
\begin{eqnarray}
\hat{\rho} &=& \alpha^{\prime}(t)|\psi(t),\downarrow \rangle \langle \downarrow ,\psi(t)| + \beta^{\prime}(t)|\psi(t),\uparrow \rangle \langle \uparrow ,\psi(t)| \nonumber\\
&& + \gamma^{\prime}(t) |\psi(t),\downarrow \rangle \langle \uparrow ,\psi(t)|
+\gamma ^{\prime *}(t)|\psi(t),\uparrow \rangle \langle \downarrow ,\psi(t)|
\nonumber\\ \label{CoherenDM}
\end{eqnarray}
The evolution of the density matrix of the system is given by Eq.\
\ref{mean field QME} with
$\hat{L}=\sqrt{\Gamma}a^{\dagger}\sigma^{-}$. Since for any operator
$\hat{A}$ the time evolution is governed by
\begin{equation}
\frac{d\langle \hat{A}\rangle}{dt}=-i\langle
[\hat{A},\hat{H}]\rangle + \frac{1}{2}\langle
[\hat{L}^{\dagger},\hat{A}]\hat{L}\rangle +\frac{1}{2}\langle
\hat{L}^{\dagger}[\hat{A},\hat{L}]\rangle \label{opdyneq}
\end{equation}
where $\langle ..\rangle \rightarrow {\rm Tr}[(..)\hat{\rho}]$, we
obtain the dynamical equations for $\psi$, $\gamma^{\prime}$ and
$\beta^{\prime}$ using Eqs.\ \ref{mean field QME} and \ref{opdyneq}.
Here $\psi = \langle\hat{a}\rangle$, $\gamma^{\prime} =
\langle\hat{\sigma}^{+}\rangle$ and $\beta^{\prime} =
\langle\hat{n}_b\rangle$ and such a procedure leads to
\begin{eqnarray}
\dot{\psi}(t) &=& i [z J \phi_a/2 - U|\psi(t)|^2\psi(t) - \lambda_0 U\psi(t)\beta^{\prime}(t) \nonumber\\
&& - \gamma^{\prime *}(t) + \mu\psi(t) ] + \Gamma \psi(t)\beta^{\prime}(t)/2 \label{psidotDSF} \\
\dot{\gamma}^{\prime}(t) &=& i [\eta z J \phi_b^*(\beta^{\prime}(t) - \alpha^{\prime}(t))/2 + \lambda_0 U|\psi(t)|^2\gamma^{\prime}(t)  \nonumber\\
&& + z V \langle n_b\rangle \gamma^{\prime}(t)/2 - \psi^*(t)(\beta^{\prime}(t) - \alpha^{\prime}(t)) - \mu \gamma^{\prime}(t)] \nonumber\\
&& - \Gamma \gamma^{\prime}(t) |\psi(t)|^2/2 \label{gammadotDSF} \\
\dot{\beta}^{\prime}(t) &=& i[\eta J z (\phi_b\gamma^{\prime}(t) -
\phi_b^*\gamma^{\prime *}(t))/2 + (\psi^*(t)\gamma^{\prime *}(t)\nonumber\\
&& - \psi(t)\gamma^{\prime}(t))] - \Gamma\beta^{\prime}(t)|\psi(t)|^2
\label{betadotDSF}
\end{eqnarray}
Numerically we obtain the steady state solutions of $|\psi|$,
$|\gamma^{\prime}|$ and $\beta^{\prime}$. We find that $\psi$ and $\gamma^{\prime}$ shows
oscillation with constant magnitude $|\psi|$ and $|\gamma^{\prime}|$
respectively and that there is a constant relative phase between
$\psi$ and $\gamma^{\prime}$. We also note from the dynamical equations that
the total density of the system, $n_{\rm total}=|\psi|^2+\beta^{\prime}$ is a
conserved quantity. Using this conserved quantity, we sketch a
typical steady state behavior of the dynamical quantities as shown
in Fig.\ref{steadySF}. From Fig.\ \ref{steadySF}(b), we see that
with increasing $\Gamma$, both the excited state density($\beta^{\prime}$)
and the excited state superfluidity($|\gamma^{\prime}|$) decrease; they
eventually vanishes for large $\Gamma$. In contrast, as shown in
Fig.\ \ref{steadySF}(a), the ground state density($|\psi|^2$)
increases and hence the ground state superfluidity($|\psi|$) also
increases with $\Gamma$. We note that this phenomenon can be
understood as follows. Since a larger number of Rydberg excitation
are destroyed for larger decay rate $\Gamma$, the ground state
density of the atoms increase so that $n_{\rm total}$ may remain
constant. In the deep superfluid limit, within the classical field
approximation the ground state number density is same as the
superfluid density $|\psi|^2$ neglecting the quantum fluctuations.
However, in the correlated regime the superfluid density
significantly deviates from the number density due to enhanced
quantum fluctuations and ground state SF order parameter decreases
with increasing $\Gamma$ as seen in the hard core limit. To study
quantum fluctuations in the ground state superfluidity, more careful
analysis is required which is beyond the scope of the simple ansatz
(Eq.\ \ref{CoherenDM}) for the density matrix.


\begin{figure}[ht]
\begin{center}
\includegraphics[scale=0.16]{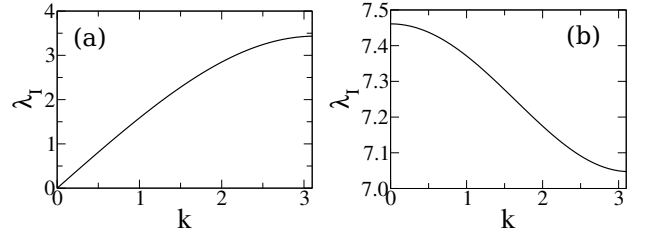}
\end{center}
\caption{(a)Plot of the dispersion of the gapless sound mode as a
function of $k$ (for diagonal cut in the Brillouin zone) in the deep
SF phase. (b) Plot of the highest gapped eigenmode as a function of
$k$. For both plots $\Gamma=0.2$. All parameters are same as in Fig.\ \ref{steadySF}}
\label{Disp}
\end{figure}

To obtain the dispersion of collective excitations in the deep SF
phase, we calculate the fluctuation above the steady state values
and decompose the fluctuation into Fourier modes similar to that for
the hardcore boson case in the strong interacting regime. Also we
choose the steady state value of $\gamma^{\prime}$ to be $\gamma_1=
\psi \gamma^{\prime} /|\psi|$ so as to have a specific relative
phase between $\psi$ and $\gamma^{\prime}$, where $\psi_1=|\psi|
\exp[i \theta_{\psi}]$ is the steady state value of $\psi$ and
$\theta_{\psi}$ is the phase of $\psi$ which is undetermined up to a
global phase factor. An analogous calculation yields
\begin{eqnarray}
\lambda \delta\psi _{\vec k} &=& i[J \epsilon(\vec k)\delta\psi _{\vec k}/2 - 2U|\psi _1|^2\delta\psi _{\vec k} - U\psi _1^2\delta \psi _{\vec k}^{*} + \tilde{\mu} \delta \psi _{\vec k} \nonumber\\
&&  - \lambda_0 U\psi _1\delta \beta^{\prime} _{\vec k}  - \lambda_0 U\beta^{\prime}
\delta \psi _{\vec k} - \delta \gamma _{\vec k}^{\prime *}] \nonumber\\
&& + \Gamma (\psi _1\delta \beta^{\prime} _{\vec k} + \beta^{\prime} \delta \psi_{\vec k})/2  \nonumber\\
\lambda \delta\beta^{\prime} _{\vec k} &=& i[\eta J
(\gamma_1^*\delta\gamma^{\prime}_{\vec k} - \gamma_1\delta\gamma^{\prime *}_{\vec
k})(z-\epsilon(k))/2 + (\psi _1^{*}\delta \gamma^{\prime *} _{\vec k} \nonumber\\
&& +\gamma _1^{*}\delta \psi _{\vec k}^{*} - \psi _1\delta \gamma^{\prime} _{\vec k} - \gamma _1\delta \psi _{\vec k})] -\Gamma(|\psi _1|^2\delta \beta^{\prime} _{\vec k} \nonumber\\
&& + \psi _1^{*}\beta^{\prime} \delta \psi _{\vec k} + \psi _1\beta^{\prime} \delta \psi _{\vec k}^{*})
\label{deefsfeq} \\
\lambda \delta \gamma^{\prime} _{\vec k} &=& i[\eta z J \gamma _1 \delta \beta^{\prime} _{\vec k} + \eta J \epsilon(\vec k)(\beta^{\prime} -1/2)\delta \gamma^{\prime} _{\vec k} - \tilde{\mu} \delta \gamma^{\prime} _{\vec k} \nonumber\\
&& - 2 \psi _1^{*} \delta \beta^{\prime} _{\vec k} - (2\beta^{\prime} -1)\delta \psi _{\vec k}^{*} + V\gamma_1\epsilon(\vec k)\delta \beta^{\prime} _{\vec k}/2 \nonumber\\
&& + z V \beta^{\prime} \delta \gamma^{\prime} _{\vec k}/2 + \lambda_0 U(\psi
_1^{*}\gamma _{1}\delta \psi _{\vec k} + \psi _1\gamma _1\delta \psi
_{\vec k}^{*}+ |\psi _1|^2  \nonumber\\ && \times \delta \gamma^{\prime}
_{\vec k})] - \Gamma (|\psi _1|^2\delta \gamma^{\prime} _{\vec k} + \psi
_1^{*}\gamma _1\delta \psi _{\vec k} + \psi _1\gamma _1\delta \psi
_{\vec k}^{*})/2 \nonumber\
\end{eqnarray}
where $\tilde{\mu} = \mu - E_f^{\prime}$ is the modified chemical
potential, $E_f^{\prime}$ is the frequency of oscillation of $\psi$
and $\gamma^{\prime}$, $\lambda$ is the complex eigenvalue,
$\epsilon(\vec k)=2\sum_{i=1..d} \cos k_i$, $d$ is the dimension and
$\delta \psi _{\vec k}$ and $\delta \gamma^{\prime} _{\vec k}$ are
the fluctuations of $\psi _1$ and $\gamma _1$ respectively. These
three equations are supported with another two equations of $\delta
\psi _{\vec k}^{*}$ and $\delta \gamma^{\prime *} _{\vec k}$. From
the resulting $5\times 5$ matrix we numerically find out the five
eigenvalues out of which one is real with $\lambda _R < 0$ and other
four complex eigenvalues come with pair(complex conjugate to each
other). Out of these two pairs of complex eigenmodes, imaginary part
of one gives gapless sound mode [see Fig.\ref{Disp}(a)] at $\vec
k=0$ while the other provides the gapped collective mode [see Fig.\
\ref{Disp}(b)]. The linear dispersion (sound mode) can be seen from
Fig.\ref{Disp}(a) and its slope yields the sound velocity $v_s$.
Similar to the superfluid density $v_s$
 also increases with $\Gamma$.

\section{ Non-equilibrium Dynamics for hardcore bosons}
\label{neqd}

In this section, we study the effect of linear ramp in $\Gamma$ and
$J$. To this end, we construct the initial density matrix of the
system from the steady states obtained in the SF phase for a given
$\Gamma_i$ and $J_i$ and linearly vary either $\Gamma$ or $J$
through its critical value. This is done, for example, by choosing
$J(t)$ to be
\begin{equation}
J(t)=J_i + (J_f - J_i)t/t_{\rm max}
\end{equation}
where $J_{f(i)}$ is the final(initial) value of the hopping matrix
element $J$ which corresponds to a steady state belonging to
the MI(SF) and $t_{\rm max}$ is the ramp time. 
A similar protocol is chosen for variation of
$\Gamma$ keeping $J$ fixed. In what follows, we concentrate on the
change in the superfluid order parameters($\alpha,\beta$) as a
function of time following and the change in the entropy(S) across
the dynamical transition. To do this, we numerically solve the
density matrix equation,
\begin{eqnarray}
\frac{\partial \hat{\rho}_i}{\partial t}&=&
-i\left[\hat{H}^{MF}_i[J(t)],\hat{\rho}_i\right]+ {\hat {\mathcal
L}} \hat \rho \label{noneq1}
\end{eqnarray}
which corresponds to variation of $J$ with fixed $\Gamma$. An
analogous equation can be easily written for the case when $\Gamma$
is varied keeping $J$ fixed.

\begin{figure}[ht]
\begin{center}
\includegraphics[scale=0.16]{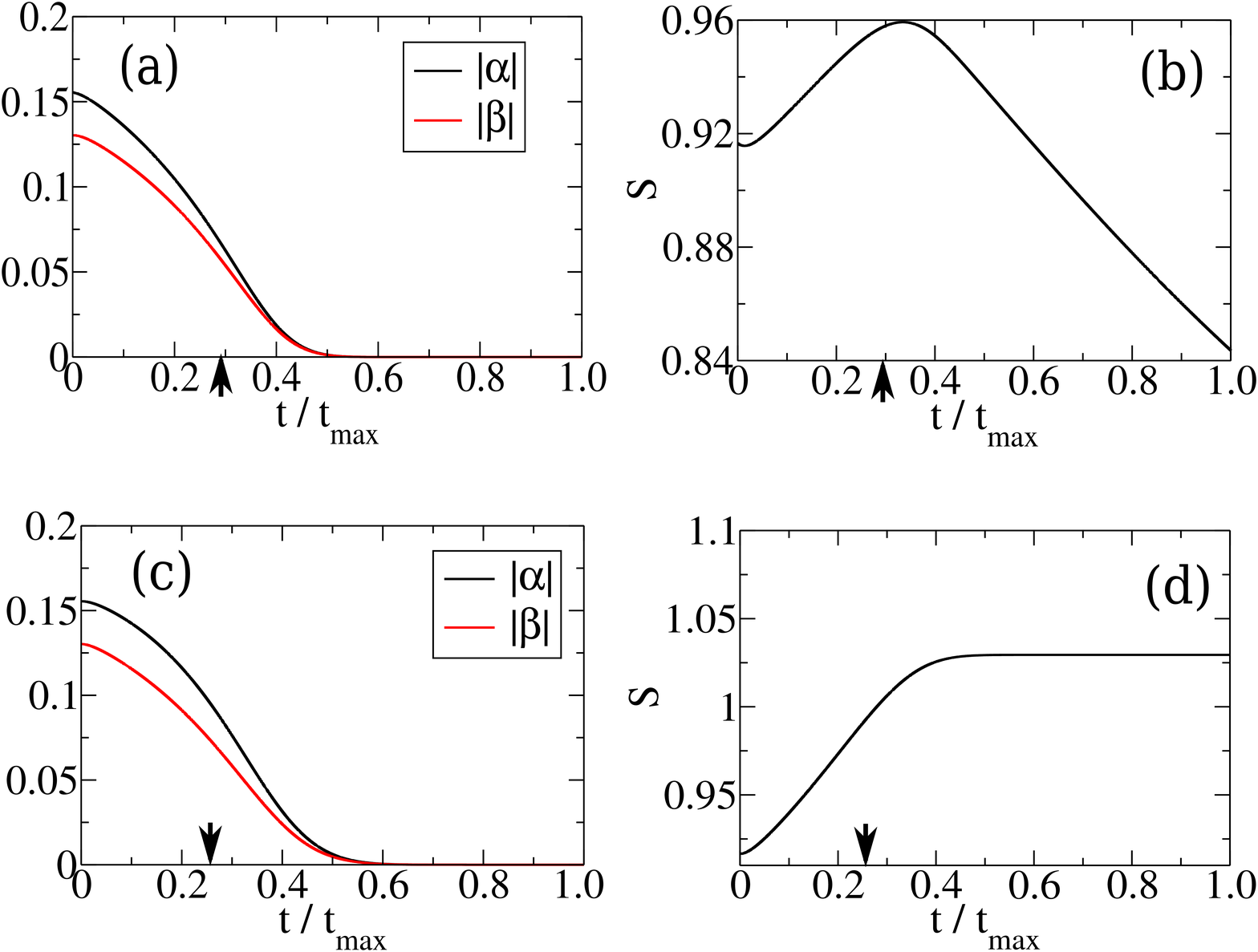}
\end{center}
\caption{(a) Plot of superfluid order parameter amplitudes
$|\alpha(t)|$ and $|\beta(t)|$ as a function of $t$ during a linear
ramp of $\Gamma$ (with $\Gamma_i=0.8$ to $\Gamma_f=2.5$ and $zJ=8$)
from the SF to the MI phase. The critical point is reached at $t=0.4
t_{\rm max}$ as marked in the figure. All other parameters are same
as in Fig.\ \ref{PhDiag1}(b). (b) Similar plot for the entropy $S$.
All parameters are same as in (a). (c) and (d) Similar plots as (a)
and (b) respectively but during linear ramp of $J$ ($zJ_i =8$ to
$zJ_f=2$) with fixed $\Gamma=0.8$. All other parameters are same as
in (a). The arrows indicate the time at which the critical point is
crossed during the dynamics. Note that $S$ shows qualitatively
distinct behavior for the two protocols.} \label{dynamics}
\end{figure}

A typical behavior of superfluid order parameters $|\alpha|,|\beta|$
and entropy $S$ as a function of time is shown in Fig.\
\ref{dynamics} for $t_{\rm max}=500$ (in units of $1/\Omega$). We
find that when $\Gamma$ is varied keeping $J$ fixed, $S(t)$
monotonically increases and peaks when the system crosses the phase
boundary (Fig.\ \ref{dynamics}(b)). Upon entering the MI phase, $S$
decreases monotonically till $t=t_{\rm max}$. In contrast, when $J$
is varied keeping $\Gamma$ fixed, the entropy $S$ monotonically
increases in the SF phase and finally saturates in the MI
phase(Fig.\ \ref{dynamics}(d)). The behavior of $S$ for the former
protocol originates from an enhanced rate of destruction of Rydberg
excitations as $\Gamma$ is increased; this leads to larger weight of
the system in the ground state and thus to vanishing $S$. In
contrast, for the second protocol where $\Gamma$ is held fixed, the
rate of decay of excitations is held constant; the dynamics merely
cease to produce new excitations due to a large Mott gap in the
low-energy sector. In addition, the fixed rate of $\Gamma$ ensures
that the system reaches a steady state. This leads to decay of $S$
in the MI phase for the first protocol and its constant value for
the second.  The difference between the behavior of $S$ for the two
protocols mentioned above can also be understood by noting that in
Eq.\ \ref{eq7}, the dynamics due to the quench of the hopping term
$J$ is generated only through its direct coupling with the SF order
parameters which decays exponentially with time in the insulating
phase; hence the effect of $J$ vanishes in the long time dynamics.
On the other hand the dissipation term $\Gamma$ is coupled to the
occupation number and other parameters which in turn gives rise to
non-trivial asymptotic behavior of $S$ for the linear quench of
$\Gamma$.

For both the cases the entropy generation rate shows a distinct
change near the phase boundary. Also, $|\alpha|$ and $|\beta|$
monotonically decrease as one moves from the SF to the MI phase
[Fig.\ref{dynamics}(a) and (c)]. This can be understood as a
characteristic of evolution of a system with finite dissipative
parameter which suppresses quantum oscillations of these quantities;
we note that this constitutes an essential qualitative difference
between time evolution of Rydberg atoms and other ultracold bosons
system where dissipation is absent.

\begin{figure}[ht]
\begin{center}
\includegraphics[scale=0.15]{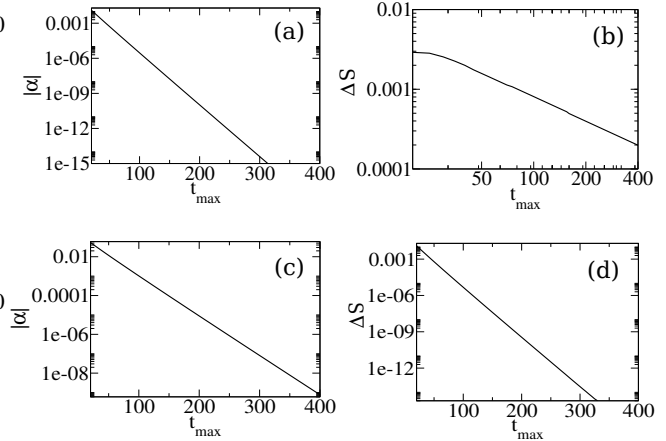}
\end{center}
\caption{(a) Plot of $|\alpha|$ and $\Delta S$ evaluated at
$t=t_{\rm max}$ (at the end of the ramp) as a function of $t_{\rm
max}$ for linear ramp of $\Gamma$ ($\Gamma_i=0.8$ $\Gamma_f=2.5$ for
$zJ=8$) are shown in log scale. $|\alpha|$ shows exponential fall
whereas $\Delta S$ shows power law fall with $t_{\rm max}$. (c)-(d)
Both $|\alpha|$ and $\Delta S$ shows exponential fall with $t_{\rm
max}$ while ramp in J from $zJ_i=8$ to $zJ_f=2$ for $\Gamma=0.8$.}
\label{dynamics2}
\end{figure}

Next, we study the deviation of entropy, $\Delta S = |S(t_{\rm max})
- S_{0}|$ from the final steady state value $S_{0}$ and the SF order
parameter amplitude $|\alpha(t_{\rm max})|\equiv |\alpha|$ as a
function of $t_{\rm max}$ for both the protocols. The variation
$|\alpha|$ and $\Delta S(t_{\rm max})$ is shown in Figs.\
\ref{dynamics2}(a) and (b) for a ramp of $\Gamma$ and in Figs.\
\ref{dynamics2}(c) and (d) for a ramp of $J$. For both the protocols
with linear ramp the SF order parameter amplitude $|\alpha|$ decays
exponentially with the ramp rate $t_{\rm max}$ which is evident from
the linear fall of $|\alpha|$ when plotted in log-scale, as shown in
Fig.\ref{dynamics2}(a) and (c). In contrast, $\Delta S$ behaves
differently for two types of linear ramping protocols. For a ramp of
$\Gamma$ with fixed $J$, we find that $\Delta S$ decays as a power
law $\sim 1/t_{\rm max}^{\kappa}$. In Fig.\ref{dynamics2}(b), we
plot $\Delta S$ as a function of $t_{\rm max}$ in log-log scale and
obtain the value of the exponent $\kappa \simeq 1.0$ from the slope
of the linear portion of the curve. In contrast for a ramp of $J$
with fixed $\Gamma$,  $\Delta S$ falls off exponentially with quench
rate $t_{\rm max}$ as shown in Fig.\ref{dynamics2}(d).

We note in this context that the behavior of defect density or
equivalently entropy as a function of ramp rate through a critical
point has been studied in several context both for closed \cite{kz,
rev} and open \cite{smitha1,patane1} quantum systems. The latter
class of system features dissipation and noise via coupling to an
external bath; this leads to a scaling of defect density $ n \sim
\gamma (k_B T)^3/v_0$, where $T$ is the temperature of the external
bath, $\gamma$ is the coefficient of dissipation, and $v_0$ is the
ramp rate. However, it is to be noted that such a scaling is derived
with the assumption that the system reaches an adiabatic regime
after crossing the critical point and that $\gamma$ remains constant
through the drive; in this case, no defect is produced or
annihilated when the system reaches the adiabatic regime for which
$\gamma, T \ll \Delta_0$, where $\Delta_0$ is the instantaneous
energy gap. In contrast,  for a ramp of $\Gamma$ for the Rydberg
atoms studied here, the system does not reach an adiabatic regime in
the MI phase once it crosses the critical point. Thus the scaling
arguments of Refs.\ \onlinecite{smitha1,patane1} can not be directly
used for explaining the scaling behavior of $\Delta S$ that we
obtain here; an appropriate analysis of this phenomenon is beyond
the scope of the present study.

\section{Discussion}
\label{dis}

In this work, we have studied non-equilibrium phases, collective
modes and quench dynamics of Rydberg atoms in an optical lattice. We
note that the spontaneous decay of Rydberg excited states implies
that a collection of Rydberg atoms constitute a non-equilibrium
quantum many body system whose dissipation may be modeled by a decay
rate $\Gamma$. In this work, we have analyzed such non-equilibrium
systems within the framework of Lindbald master equation describing
the time evolution of the density matrix within mean-field theory.

In the first part of this work, we have considered the Rydberg atoms
in the limit of on-site repulsion between the atoms and obtained the
fixed points of their density matrix equations corresponding  to
both the SF and MI steady states. In contrast to the usual MI phases
of ultracold atoms systems, we find that the present system allows
for insulating phases that can have incommensurate filling. We
analyze the fixed points and find that MI-SF transition in these
systems are continuous which may be inferred from the discontinuous
jump of $dS/d\Gamma$ at the transition. We have also numerically
studied the steady states of the Rydberg atoms with finite on-site
repulsion by constructing the initial density matrix from the
Gutzwillers wavefunction. We find that the transition in such
systems are qualitatively similar to the one for atoms with infinite
on-site repulsion. This justifies our analysis for the hard core
bosons from the point of view of possible experiments on the system.

We have also looked into the collective modes of such systems by
carrying out a linear stability analysis of the MI and SF fixed
points. The real part of the corresponding eigenvalues indicates
stability of the corresponding phases, whereas the imaginary part
give collective excitation frequencies. Such an analysis yields both
gapless sound mode and gapped collective modes in the superfluid
phase; in contrast, all modes in the insulating phase are gapped.
The velocity of the sound mode vanishes at the phase boundary; also
the energy gap increases linearly in the insulating phase near the
transition point. This allows us to infer $z=2$ and $\nu=1/2$ for
the MI-SF transition within mean-field theory.

We have analyzed the Rydberg system in the deep SF limit for weak
$U$ by using a coherent state representation of the system density
matrix. Our analysis results in a dynamical equation of the
condensate in this limit; this equations is the counterpart of the
Gross-Pitaevskii equation for dissipative systems. We analyze this
equation to find that in the classical regime where the boson
fluctuations can be ignored, both the ground state density and the
superfluid order parameter amplitude increases with increasing
dissipation strength. This, somewhat counterintuitive, result
originates from the fact that the total system density is a constant
of motion in this regime; thus a large $\Gamma$ which leads to
larger loss of Rydberg excitations automatically leads to a larger
ground state density. We also obtain the gapless sound mode and the
gapped collective modes by a stability analysis of the density
matrix equations around the SF fixed point in this regime.

Finally we investigated the non-equilibrium dynamics for a linear
ramp of either $J$ or $\Gamma$ which takes the system from the SF to
the MI phase. Unlike systems without dissipation, for linear ramp,
the superfluid order parameters do not show any oscillations and
decay monotonically in the insulating phase due to the presence of
dissipation for both the protocols. In contrast, we find that the
behavior of the entropy $S$ as a function of time during the linear
ramp is qualitatively different for a ramp of $\Gamma$ with fixed
$J$ and for a ramp of $J$ with fixed $\Gamma$; for the former
protocol, $S$ decreases with time while it remains constant for the
latter protocol once the system enters the MI phase. However, in
both cases $S$ changes slope near the phase boundary. We also
investigate in the long time dynamics the deviation of the entropy
from its steady state value, which shows exponential decrease with
ramping rate for the linear quench of $J$. In contrast, $\Delta S
\sim 1/t_{\rm  max}$ for the quench of $\Gamma$. Such a power law
arises out of the reduction of the density of Rydberg excitation of
the system with increasing $\Gamma$; its detailed analysis is left
for a possible future project.

Several recent experiments has been carried out for systems of
Rydberg atoms \cite{exp1, exp2, exp3}. The easiest experimental
verification of our result would constitute performing boson
momentum  distribution measurement for a system of Rydberg atoms in
a 3D optical lattice. Such measurement would detect the emergence of
a central peak in the momentum distribution, as measured in a
standard time of flight experiment \cite{bloch1}, as one enters the
SF phase. Thus such measurements may be used to chart out the MI-SF
phase boundary in the $J-\Gamma$ plane. The corresponding collective
modes can also be measured by standard lattice modulation, RF, or
Bragg spectroscopy in these systems \cite{bragg1}.

In conclusion, we have studied a system of Rydberg atoms in the
presence of dissipation using a density matrix formulation within
mean-field theory. Our work constitutes an analysis of these
dissipative atom system which may take into account their SF phases
by going beyond the frozen atom limit; it may thus serve as the
first step towards a more complete picture of the description of the
phases and dynamics of these atoms beyond mean-field theory where
the effects associated with quantum fluctuations are treated more
rigorously. We have charted out the MI-SF phase-boundary of such a
system and derived expressions for the collective modes in each of
the phases. We have also studied non-equilibrium dynamics of these
systems for linear ramp protocol of both $J$ and $\Gamma$ and
identified a $1/t_{\rm max}$ scaling of $\Delta S$ for ramp of
$\Gamma$. We have suggested experiments which may test our theory.

\end{document}